# Ultrabroadband photosensitivity from visible to terahertz at room temperature

Dong Wu[1]*, Y. C. Ma[2], Y. Y. Niu[3], Q. M. Liu[1], Tao Dong[1], S. J. Zhang[1], J. S. Niu[1], H. B. Zhou[1], Jian Wei[1], Y. X. Wang[3], Z. R. Zhao[3], N. L. Wang[1]*

[1]International Center for Quantum Materials, School of Physics, Peking University, Beijing 100871, China

[2]School of Materials Science and Engineering, Tianjin University of Technology, Tianjin 300384, China

[3]Key Laboratory of Particle & Radiation Imaging, Department of Engineering Physics, Tsinghua University, Beijing, 100084, China

*Correspondence to: dongwu@pku.edu.cn; nlwang@pku.edu.cn.

**Abstract**

Charge-density wave (CDW) is one of the most fundamental quantum phenomena in solids. Different from ordinary metals in which only single particle excitations exist, CDW also has collective excitations and can carry electric current in a collective fashion. Manipulating this collective condensation for applications has long been a goal in the condensed matter and materials community. Here we show that the CDW system of 1T-TaS$_2$ is highly sensitive to light directly from visible down to terahertz, with current responsivities around the order of ~1 AW$^{-1}$ at room temperature. Our findings open a new avenue for realizing uncooled, ultrabroadband and sensitive photoelectronics continuously down to terahertz spectral range.

**Introduction**

The ability to convert light into electronic signal is critical in photoelectronics. The effect has broad applications ranging from imaging, communication, quantum information to space science. However, high sensitivity to light over broad spectral range, especially in the low frequencies down to terahertz (THz) at room temperature (RT), is particularly rare but meritorious. Conventional semiconductors such as silicon gain the photoresponse properties through the single-particle excitations across the band gap and are transparent to the energy below the gap to THz photons *(1-4)*. On the other hand, the photoresponse properties of simple metals or very-narrow-gap semiconductors suffer seriously from free-carriers screening and fast-quenching effect, especially at elevated temperatures *(4,5)*. Solids with collective electronic states *(6,7)*, on the contrary, can have electronic response in a collective fashion and behave quite differently from the single-particle excitations.

Charge-density wave (CDW) is one of the most studied collective electronic states in solids. Different from ordinary metals in which only single particle excitations exist, CDW materials also have collective excitations being referred to as an amplitude mode and a phase mode. Amplitude excitation behaves as optical phonons and is not expected to have direct effect on the electric transport properties, while phase excitation corresponds to the translational motion of the CDW condensate and can has dramatic effect on charge transport properties. The phase mode is usually pinned at finite frequency. By applying dc electric field, the phase mode can be driven into a current-carrying state being usually referred to as sliding CDW, leading to nonlinear current-voltage (I-V) characteristics *(6,7)*. As a result, the frequency- and electric-field- dependent conduction exhibits characteristic features of the translational motion of the CDW condensate. Here we show that incident illuminations can induce significant changes in electronic properties in 1T-TaS$_2$, a well-known two-dimensional (2D) CDW compound, making it a superior system for realizing high sensitive photoelectronics directly in an ultrabroadband range from visible (Vis) to infrared and even THz, at RT.

1T-TaS$_2$ has a crystalline structure consisting of planes of hexagonally arranged tantalum atoms sandwiched by two layers of sulphur atoms, as shown in Figure 1A. As one of the most studied CDW compounds of 2D transition-metal chalcogenides, 1T-TaS$_2$ hosts multiple equilibrium states arising from the interplay between electron correlation, lattice strain and Fermi surface nesting *(8-13)*. Upon cooling across ~550 K, the system transfers from a simple metal into an incommensurate (IC-) CDW state with an associated modulations of lattice distortions. The distortions sharpen to form star-shaped polaron clusters (See inset of Fig. 1B) and become nearly commensurate (NC-) below $T_c$ = 350 K, being referred to as the phase of NC-CDW. Further cooling leads to another first-order phase transition near 183 K, below which the system becomes a gapped commensurate (C-) CDW state. These variations have made 1T-TaS$_2$ an ideal playground for investigating CDW dynamics *(12,13)* and novel metastable states*(14,15)* via external manipulations.

**Results**

High-quality single crystals of 1T-TaS$_2$ were synthesized by chemical vapour transport method in a sealed quartz tube with a subsequent quenching process to retain the 1T phase (see Fig. S1-S3 for more information). Figure 1B shows the temperature dependent resistivity of our 1T-TaS$_2$ sample which captures the CDW phases upon cooling from 400 K. Above the critical temperature $T_{c1}$ = 350 K, the resistivity keeps almost a constant value. Then it undergoes an abrupt jump at $T_{c1}$, reflecting a phase transition from IC-CDW to NC-CDW *(13,15)*. Near RT, the resistivity values have the order of 1 mΩ·cm, where the values are close to the maximum universal values predicted in the 2D metals *(16,17)*. Different from the IC-CDW phase, the ρ-T curve in NC-CDW phase has an abnormal negative slope. Upon further cooling down to 183 K, the resistivity shows another abrupt upward jump, corresponding to the formation of the C-CDW state. A significant hysteresis is seen between cooling down and warming up the sample, reflecting the first-order phase transition.

Two-terminal devices are used to investigate the photo-active electric properties of the NC-CDW phase of 1T-TaS$_2$ at RT. The prototype devices are fabricated on exfoliated thin flakes in thickness of several micrometers with the gold electrodes prepared by deposition method. Then the flakes are cut into slim channels and transferred onto sapphire substrates for using as devices. The resultant structure is depicted in Figure 1C. A 1T-TaS$_2$ device with the channel (length 780 μm, width 23 μm) between the electrodes is used for illustrating our results. The representative current-voltage (I-V) curves measured at RT under dark are plotted in Figure 1D for both sweeping up and down modes. Both curves can be divided into three different intervals: a linear section at low electric fields, a non-linear section followed by a current jump-up, and a high conduction section thereafter. The conductivity value at high-field ends is close to that measured in IC-CDW state. Previously, it has been revealed for a field-driven switching in I-V (below 180 K) characterizing a phase transition from C- to IC- CDW in 1T-TaS$_2$ system *(13)*. By analogous considerations, the observed switching in I-V at RT might represent the phase transition from NC- to IC- CDW. A more plausible mechanism responsible for this switching character is relating with CDW sliding dynamics driven by the applied electric field *(6,8)*. Below the sliding threshold voltage $V_T$, the CDW condensate is primarily pinned to the background lattice. While near $V_T$, a significant non-linear I-V interval emerges, corresponding well to the depinning of CDW condensate transport. Such electric-field sensitive characteristics have been widely observed and investigated in traditional CDW materials, such as $NbSe_3$, $TaS_3$ and molybdenum blue bronzes *(18-21)*. As a result, in $1T-TaS_2$ device the conduction phase changing between low-conduction (LC) and high-conduction (HC) states is manifested (Fig. 1D).

Unlike usual semiconductor which is transparent below the band energy gap, $1T-TaS_2$ crystal has rather high in-plane reflectivity ($R$) at low energy. Figure 2A shows the value of 1-$R$ as a function of wavelength from Visible down to THz region at RT, which roughly reflects the absorption of the crystal sample. The penetration depths are estimated to be in the range between ~40 nm (at $\lambda=1$ μm) in nir-infrared (Nir) and ~0.6 μm (at $\lambda=100$ μm) in THz at RT respectively (see materials and methods). When illuminating, the I-V function of $1T-TaS_2$ devices was found to be reshaped. Figure 2B depicts such a typical current response for a laser illumination of $\lambda = 1550$ nm. Seen clearly in the figure, the whole I-V curve under illumination is lifted above the dark one. The current response is relatively weak at low biases, but it gets more and more drastic as the applied biases increasing into the nonlinear interval approaching the switching edge. These effects can be characterized to be the direct photoresponse of $1T-TaS_2$ material, without competitive influence from the contacts. The current response increases with light-intensity increasing, accompanied by a gradual reduction of the transition edges (see Fig. S4 for more information). The threshold $V_T$ exhibits a quasi-linear relationship with the incident intensity (Fig.2C), suggesting that the CDW pinning force can be tuned by the illumination fluences.

Figure 2D illustrates the current switching effects with a constant illuminating intensity is turn on and off. Away from the threshold voltage (for instance at the bias voltage of 0.71 V, Fig. 2D, left panel), stable photocurrent switching is recorded as the illuminating is modulated by a shutter. When the light-intensity is high enough, it will drive the shifted switching edge completely into the fields below the initial threshold voltage $V_T$, giving rise to a bias window connecting the LC with HC conduction states (see Fig. 2B). Comparatively, for the applied potential locating inside the bias window, e.g. at 0.73 V, large photoeffect corresponding to the conduction state transition from LC to HC state could be induced (Fig. 2D, right panel), indicating that photoexcitation could be an efficient method to control the dynamical CDW phases in $1T-TaS_2$.

For a given applied voltage below $V_T$, upon increasing the incident intensity, the current also shows a critical intensity $P_c$ in $I$-$P$ function (where $P$ is the incident intensity), as shown in Fig. 3A. Below $P_c$, the corresponding $I$-$P$ curve demonstrates a quasi-linear behavior. The range of the linear photoresponse region depends strongly on the applied bias. As the intensity increases, the current response jumps to a higher value corresponding to a higher constant conductance, indicating that the LCS-HCS state transition is triggered finally.

Surprisingly ultrabroadband photosensitivity, similar to aforementioned photoresponse effects, are observed at wavelengths ranging from visible down to terahertz at RT in $1T-TaS_2$ device (See Fig. S6 for more information). As for intensive illuminations, the LC-HC conduction phase transition can be induced. While for weak illuminations, the I-P function keeps in a quasi-linear behavior that the ratio of $I_p / P$ manifests independent on the incident power, where $I_p$ is the net photocurrent, being defined as the photoexcited current minus the dark current. We numerically investigate such a nominal current responsivity, $R_{fm} = I_p / P$, where the $R_{fm}$ figure-of-merit quantifies the signal component of the sensitivity and is important to photodetection applications *(5)*. As shown in Figure.3B, high current responsivities are recorded by adjusting the applied potentials approaching $V_T$, with the magnitude between 0.25 and 3.92 AW$^{-1}$ being obtained at bias voltage 0.72 V for an instance (See Fig. S7 for more information). At THz wavelength $\lambda = 118.8$ μm, the responsivity is 0.76 AW$^{-1}$, which is about two orders higher than that detected in graphene based ultrabroadband and THz detector systems *(22)* (see Table. S1 for more information). To the best of our knowledge, few materials were reported to demonstrate potentials for such the direct ultrabroadband down to THz photoresponse, as referring to graphene, topological insulators and recently a telluride of EuSbTe$_3$ *(22-24)*. Compared with them, 1T-TaS$_2$ is the system presenting exceptionally strong photoresponse, which is likely to be related to the complex CDW dynamics, showing significant potentials of RT photodetection across the wavelengths range continuously from visible to THz regimes.

To facilitate the applications, the noise spectra of the device as a function of frequency is analysed *(25)*. It manifests a structureless broadband noise in $1/f^\alpha$ at low frequencies, which approaches the Johnson–Nyquist limit at higher frequencies. We calculate the noise equivalent power (NEP), i.e. the signal power when signal-to-noise ratio is unity *(5)*. Operating basically at electric-fields with linear photoresponse, the named NEP is evaluated to be ~ 80 pW·Hz$^{-1/2}$ at $\lambda$ = 532 nm for the applied bias 0.71 V (See Fig. S8 for more information). At THz frequencies, the available NEP ~ 0.4 nW·Hz$^{-1/2}$ (at $\lambda$ = 118.8 μm), rivals the state-of-art RT THz detectors such as thermal bolometers and Golay cells *(26)*.

It is important to characterize the time scale of photoresponse effect. It appears that the intrinsic photoresponse time scale cannot be correctly measured by using a normal shutter or mechanical chopper with low frequency modulation since the photoresponse is too fast. We employed a pulse femtosecond laser using a regenerate amplifier and optical parametric amplifier (OPA) with a repetition rate of 1 KHz to investigate the photoresponse of 1T-TaS$_2$ devices at RT. The temporal 'time-of-flight' response *(27,28)* by the photo-excitations under applied bias voltage is recorded by an oscilloscope with a bandwidth of 350 MHz, for which a channel in 50 nm thickness and 30 μm length between the electrodes is used for measurement. The result is shown in Fig. 4. The pulsed current signal rises up immediately after the photo-excitation. The rising time is recorded to be ~1.5 ns, which is still limited by the time resolution of the oscilloscope electronics. The photoexcited signal shows approximately a plateau below 30 ns, then falls fast in another 30 ns followed by a rather slow tail. The rather long-lasting period of the plateau indicates that the sample is driven to a metastable state, which can be considered as a local variation of the CDW phase *(14,15,29)*. The slow tail is found to be bias-dependent, which might be related to the defects and external stress *(15)*. It deserves to remark that the photoresponse can be hardly detected when the applied bias exceeds the threshold voltage $V_T$ (See Fig. S9 for more information). The results suggest that the experimental observations can be attributed to the NC-CDW dynamical photoresponse.

**Conclusions**
To summarize, we have demonstrated that the layered 2D CDW system 1T-TaS$_2$ is highly sensitive to light at RT with ultrabroadband photoresponse from Vis to THz spectral range. The revealed characters have direct relations with the inherent collective electronic dynamics of the CDW state and the results may extend well beyond the 1T-TaS$_2$ material case. The advantages of uncooled and ultrabroadband photoresponse make this two-dimensional chalcogenide highly attractive for exploring more efficient photoelectronics, such as novel memory devices, photodetectors and spectroscopy, from both experimental and theoretical perspectives.

## Materials and Methods

Samples preparation:

1T-TaS$_2$ single crystals were grown by the chemical vapor transport (CVT) method with iodine as a transport agent. The high-purity Ta (4N) and S (4N) were mixed in chemical stoichiometry with additional iodine (~2% total mass) and heated at 850°C for 2 days in an evacuated quartz tube. Then the quartz tube was transferred to a two-zone furnace, where the source zone and growth zone were fixed at 880°C and 780°C for two weeks. The tube was then quenched in cold water to ensure retaining of the 1T phase. Large shinning crystal plates with common size of ~ 3×3 mm$^2$ can be yielded.

For devices preparing, thin flakes were exfoliated from the bulk crystals by means of scotch tapes method. Two terminal electrodes were prepared with gold deposition method on the flakes. The flakes were cut into slim bars for preparing the devices. The devices with different dimensions are checked for photoresponse properties, and they show analogous response features.

Optical spectral properties:

For optical spectra analysis, the near-normal incident reflectance spectra were measured on freshly cleaved 1T-TaS2 bulk plates (thickness ~ 50 μm), by a Bruker 80 v/S spectrometer in the 50 to 24000 cm$^{-1}$ frequency range. An *in-situ* overcoating technique was employed for reflectance measurements *(30)*. The penetration depth δ$_0$ is estimated by $1/δ_0 = 4π ω k /c$, where k is the extinction coefficient which is available from the reflectivity spectrum by Kramers-Kronig transformation. The penetration depths are evaluated to be far smaller than 0.5 μm from UV to THz at room temperature (for instance, it is ~ 45 nm at λ = 1 μm and ~ 0.6 μm at λ = 100 μm respectively). Thus the absorbance spectrum applied to the thick samples (i.e. several micrometers) can be roughly evaluated by α = 1-R, where α is the absorbance and *R* the reflectance.

Electrical and photoresponse properties:

The DC electrical signals are measured using Keithley 2602B sourcemeters. For photocurrent measurements, various continuous-wave lasers are provided for light sources, as referred to the solid-state lasers (λ = 532 nm, 635 nm, 1064 nm,1550 nm), MIR source (10 μm, CO$_2$ laser). A far-infrared gas laser (FIRL 100, Edinburgh Instruments Ltd.) is used serving as the THz illumination source (163 μm, 118.8 μm, 96.5 μm). The incident power was monitored by calibrated power meters. The continuous-wave lasers combining with a shutter were provided for the tests of current switching effect.

The absolute responsivity is calculated by:
$$R_{fm} = (I_{on} - I_{dark}) / P$$
Where $R_{fm}$ is in unit of AW$^{-1}$, $I_{on}$ is the current under illumination, $I_{dark}$ is the dark current and *P* the incident light power,

The NEP is estimated by NEP = $I_n / R_{fm}$, where $I_n$ is the noise current in A·Hz$^{-1/2}$, $R_{fm}$ the responsivity in unit of AW$^{-1}$, and NEP in unit of W·Hz$^{-1/2}$. For evaluating NEP values, the noise current is extracted from the noise power spectra at frequency bandwidth of 50 Hz, and the photoresponsivities are estimated at the same frequency.

The noise power spectra are measured using home-made amplifiers and then digitized with a data acquisition card. A cross correlation algorithm is used to average out the instrument noise. Just battery and ballast resistors were used to avoid extra noises due to external circuit. The device is kept in a shielded dark enclosure.

For temporal photoresponse analyses, pulsed femtosecond lasers at wavelengths λ = 800 nm (100 fs, 1.2 mW·cm$^{-2}$) and λ = 2.5 μm (150 fs, 1 mW·cm$^{-2}$) with the repetition rate of 1 KHz are provided as laser sources. The temporal electrical signals are recorded by an oscilloscope (Agilent Technologies DSO-X 3034A, the bandwidth is 350 MHz).

**Funding:** This work was supported by the National Science Foundation of China (No.11327806,GZ1123), the National Key Research and Development Program of China (No.2016YA0300902, 2017YFA0302904).

**Figures and Tables**

# Figure 1

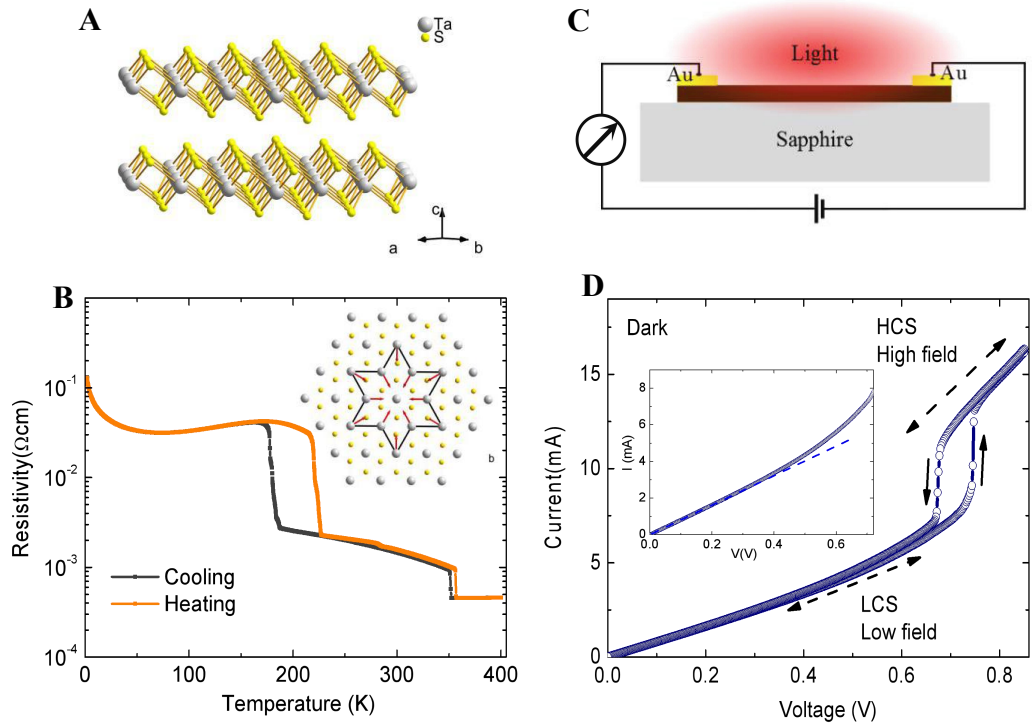

**Fig. 1. Room temperature Near commensurate CDW phase in 1T-TaS$_2$ and its electronic properties.** (**A**) Schematic images of layered structure. (**B**) Temperature dependence of the four-probe resistivity on temperature cycling. The inset depicts the lattice distortions associated with nearly commensurate CDW phase. (**C**) Schematic of the two terminal device used for photoresponse characterization. (D) The current-voltage curve under dark for the two terminal device measured under voltage sweep up and down modes at RT. The abrupt switching between low conduction state and high conduction states can be driven by the applied bias. The inset shows the quasi-linear I-V at low fields (the blue dashed is a guideline), and a nonlinear behavior at the relative high fields.

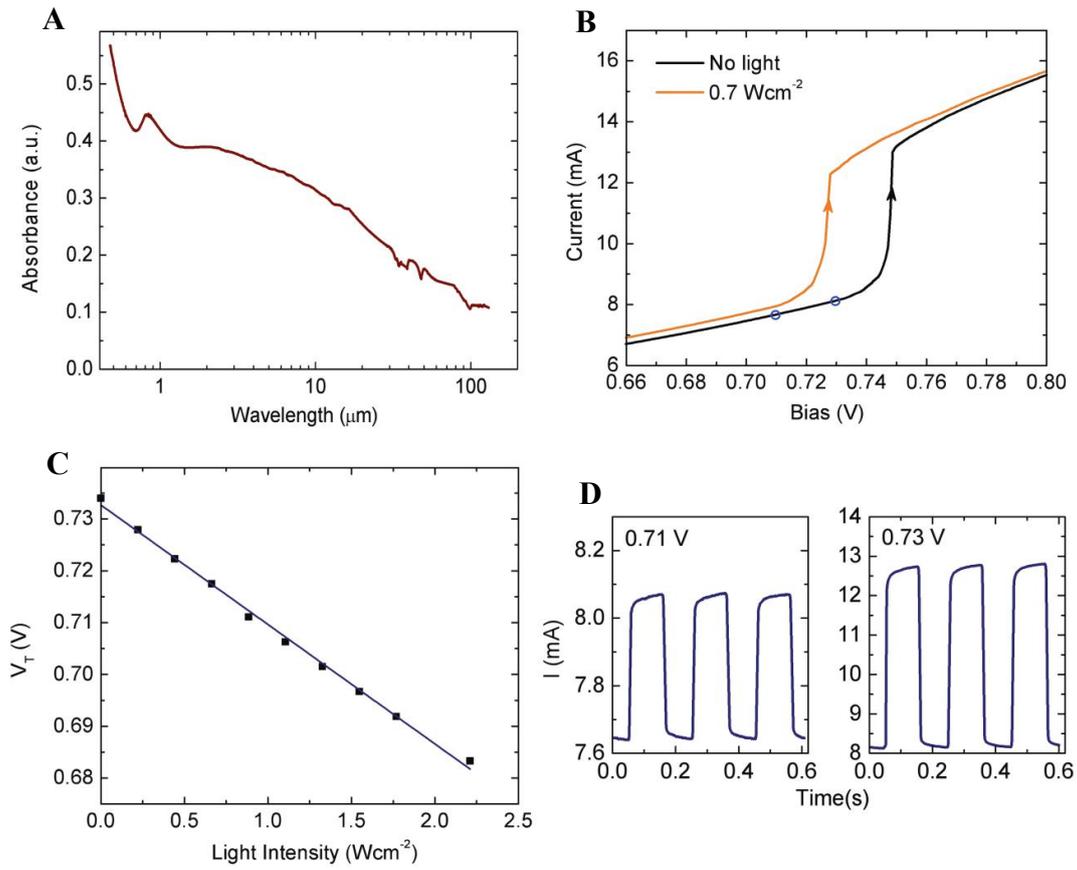

**Fig. 2. Absorbance spectra and electrical response under illuminations**. (**A**) In-plane optical absorbance spectrum of freshly cleaved surface of 1T-TaS$_2$ crystal samples. (**B**) The photoresponse of dc electrical current of a device for voltage sweep up mode. The light-intensity is 0.7 W·cm$^{-2}$. (**C**) The threshold voltage V$_T$ changes linearly with the incident light intensity. (**D**) The current switching effects for applied bias 0.71 V (left panel) and 0.73V (right panel), as bias values are marked out as blue circles respectively in (B). For the experiment, the illumination (0.7 W·cm$^{-2}$) is chopped at a frequency 5 Hz. The illuminations were provided by a continuous-wave solid-state laser at λ = 1550 nm with a laser spot in diameter ~2.7 mm focusing on the device.

**Figure 3**

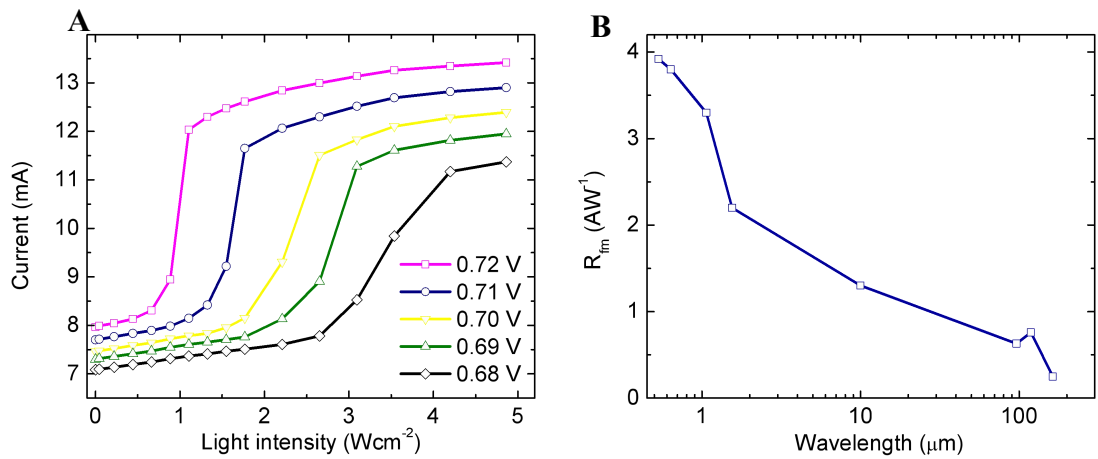

**Fig. 3. The ultrabroadband photoresponse.** (A) The current response as the function of incident intensities was investigated at illumination of λ = 1550 nm for applied voltages below the initial threshold of dark. For relative weak illuminations, a linear relationship for the current with incident intensity can be established. While, for intense illuminations, the conduction transition of LCS-HCS achieved. (B) The ultrabroadband photoresponsivity measured at external bias 0.72 V. The illuminations were provided by continuous-wave solid-state lasers from Vis to THz at λ= 532 nm, 635 nm, 1064 nm, 1550 nm and λ= 10 μm, 96.3 μm, 118.8 μm, 163 μm, respectively (See materials and methods).

**Figure 4**

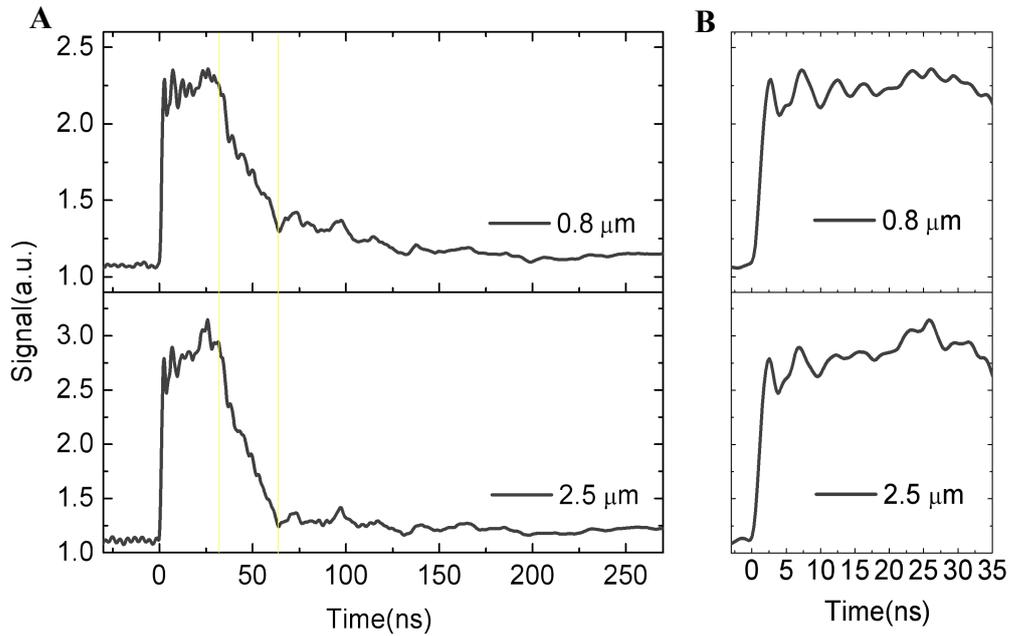

**Fig. 4: Temporal photoresponse studied by pulse excitations.** (A), Rise and fall edges showing fast and slow components. The yellow lines are guides identifying three different relaxation processes including a short retention ~ 30 ns, a fast fall in another 30 ns, and a slow tail. The small superimposed oscillations are due to ringing of the circuit. (B), Close ups of the fast components of the rise edges and short retention. The rise time shorter than 2 ns can be recognized, the accurate rise time analysis is out of the limitation of our oscilloscope. Here in the experiments, the pulsed femtosecond lasers are provided at wavelengths $\lambda$ = 0.8 μm (~100 fs, 1.2 mJ·cm$^{-2}$) and $\lambda$ = 2.5 μm (~150 fs, 1 mJ·cm$^{-2}$) with the repetition rate of 1 KHz.

## Supplementary Materials

**Materials characterization.**

**EDS analysis:** The energy dispersive spectrum (EDS) analysis of the as-grown 1T-TaS$_2$ is shown in Figure S1.

**Infrared spectral analysis:** The infrared spectra were measured on freshly cleaved 1T-TaS2 bulk plates using near-normal incident reflectance method by a Bruker 113 spectrometer in the 30 to 360 cm-1 frequency range. An *in-situ* overcoating technique was employed for reflectance measurements *(30)*. We investigate typically the infrared activated phonons of 1T-TaS$_2$ at temperatures of 300K and 78K. At 300K, the spectra of the modes are heavily suppressed by free electron excitations. As cooling to 78K, the electrons get mostly localized and the modes are clearly observed at 54, 69 79, 99, 104, 107, 120, 207, 241, 257, 261, 287, 292, 306 cm$^{-1}$. These phonon modes are separated into two groups, Fig.S2. The first group is situated at frequencies below 130 cm$^{-1}$ and the second group at the frequencies above 190 cm$^{-1}$. Because Ta atoms are about five times heavier than S atoms, the low-frequency group can be assigned to Ta vibrations and the upper to S vibrations, as reported by the reference *(31)*.

| Element | Weight% | Atomic% |
|---|---|---|
| Ta L | 74.53 | 34.15 |
| S K | 25.47 | 65.85 |
| Totals | 100 | |

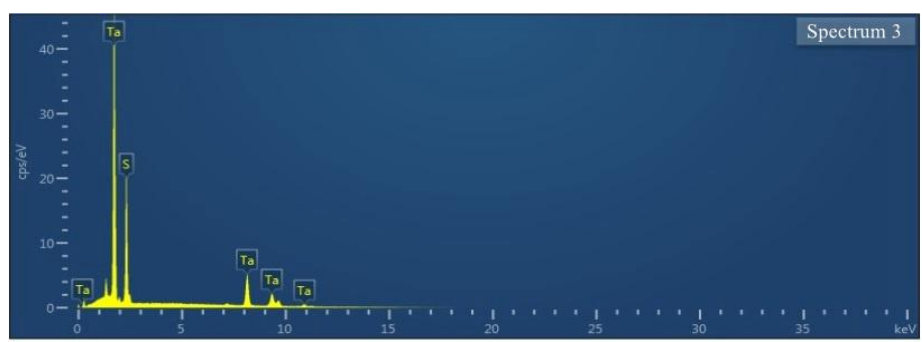

**Fig.S1**. EDS analysis of the 1T-TaS$_2$ as-grown crystals.

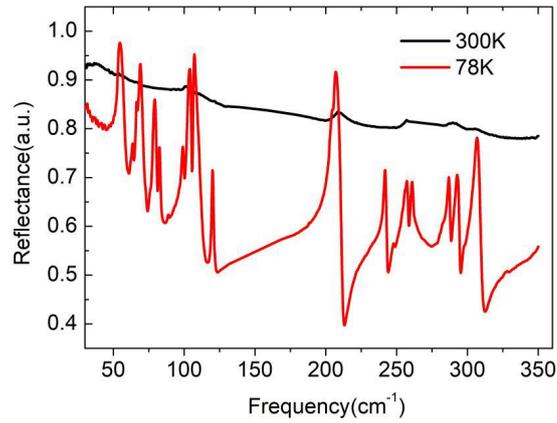

**Fig.S2**. In-plane infrared spectra of the 1T-TaS$_2$ measured at 300 K and 78 K.

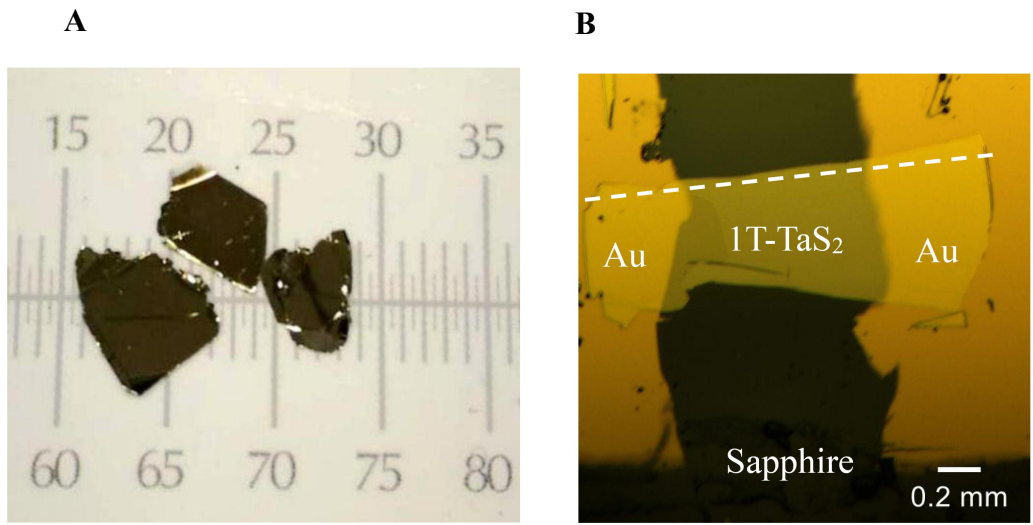

**Fig.S3**. (A), Optical images of 1T-TaS$_2$ as-growth crystals, the scale is in millimetres. (B), Schematic depiction of the device preparing. Thin flakes were exfoliated from the bulk crystal plates by means of scotch tapes method, then transferred on the sapphire substrate. Two terminal electrodes were prepared by gold deposition. The flakes were cut into slim bars before used for photoresponse experiments. The white dashed is guide for the cutting direction.

**The photoresponse spectra as the function of applied bias and incident-intensity.**

The DC current were measured using a Keithley 2602B sourcemeters. Illumination sources were provided by various continuous-wave lasers as referred to the solid-state lasers ($\lambda$ = 532 nm, 635 nm, 1064 nm,1550 nm), a mir-infrared source (10 μm, $CO_2$ laser), and a far-infrared gas laser (FIRL 100, Edinburgh Instruments Ltd.) serving as the THz illumination sources ($\lambda$ = 163 μm, 118.8 μm, 96.5 μm). The incident power of the laser beam was monitored by calibrated power meters. The continuous-wave lasers combining with a shutter were provided for the measurement of current switching effect.

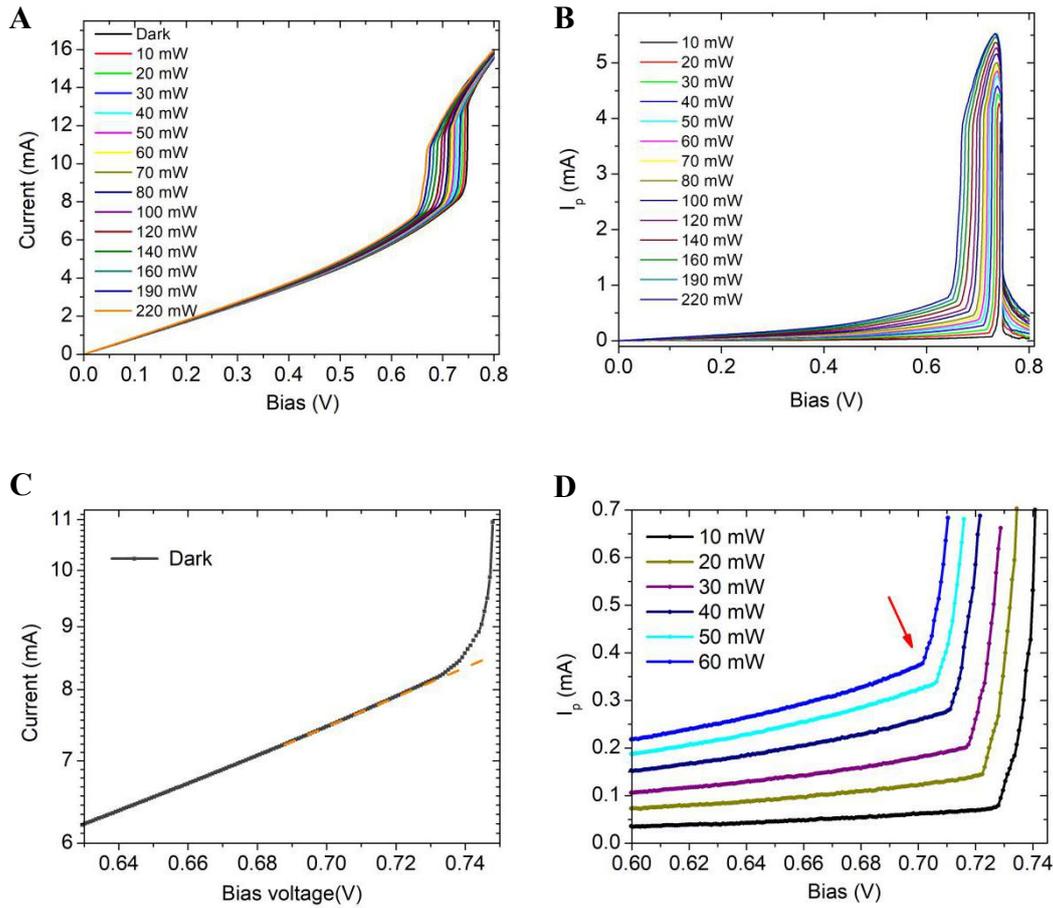

**Fig. S4:** I-V characteristics of the device as the function of applied bias and incident-intensity. (A), I-V characteristics of the device as the function of the incident-intensity and electric-field. The illuminations were provided by a continuous-wave solid-state laser at $\lambda$ = 1550 nm with a laser spot in diameter ~2.7 mm focusing on the device. The total power of laser beams was calibrated by power meters and shown in the figures. (B), The net photo-induced current $I_p$, by $I_p = I_{on} - I_{dark}$, where $I_{on}$ is the current under illuminations and $I_{dark}$ is the dark current. (C) and (D) illustrate the threshold voltage under dark and illuminating cases, respectively.

**Estimation of the temperature rise.**

After a constant light incident, the steady state value of temperature rise $\Delta T$ can be estimated using Fourier's law of heat conduction.

$$p = -\frac{k \cdot A}{d}\Delta T$$

Where the $P$ is the heat flux per unit time, $k$ is the material's conductivity, $A$ the contact area for each fluid side, $d$ the heat transfer length, and $\Delta T$ is the temperature gradient between object and environment. For 1T-TaS$_2$, the out-plane thermal conductivity $K_c$ is set to be ~ 0.01 W/(cm·K) *(14)*, which is assumed as ten times smaller than the measured in-plane $K_{ab}$ ~ 0.1 W/(cm·K) *(32)*. To our sample with the dimension of ~ 780×23×6 µm, the temperature rise $\Delta T$ is thus calculated to be less than 0.05 K for the illuminating intensity 1W/cm$^2$ at various wavelengths in our measurement range.

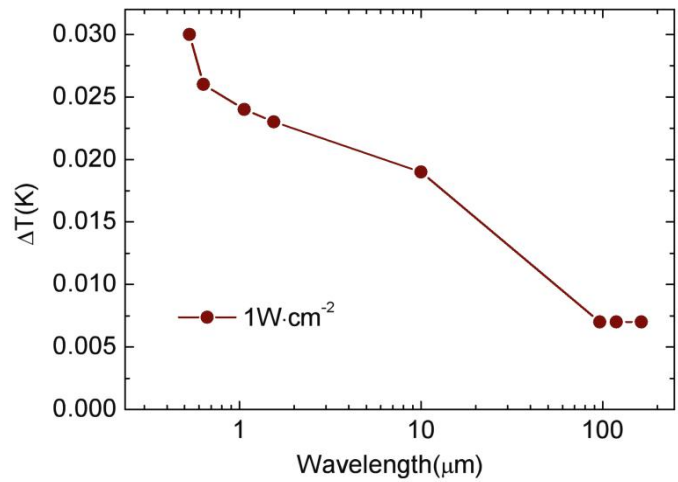

**Fig. S5:** The estimated steady state value of temperature rises for the illumination ~1 W·cm$^{-2}$ at various wavelengths in our measurement range for a sample with the dimension ~ 780×23×6 µm.

**THz response characterization:**

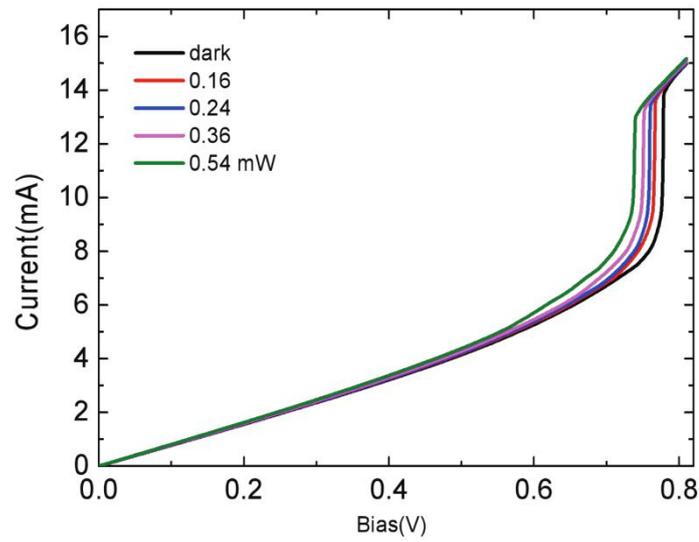

**Fig. S6:** THz response characterizations at λ=118.8 μm. A far-infrared gas laser (continuous-wave) is used serving as the THz illumination source. The laser spot diameter is ~1.6 mm. The integrated excitation power over the surface of the device are shown.

**The ultrabroadband photoresponsivity operating at quasi-linear photoresponse conditions**

The absolute responsivity is calculated by:
$$R_{fm} = (I_{on} - I_{dark}) / P$$
Where $R_{fm}$ is in unit of AW$^{-1}$, $I_{on}$ is the current under illumination, $I_{dark}$ is the dark current and $P$ the incident light power,

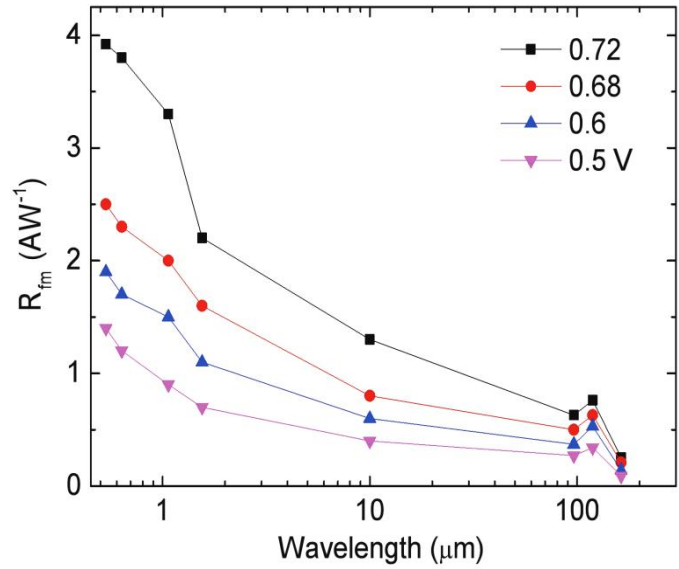

**Fig. S7:** The ultrabroadband photosensitivity operating at quasi-linear photoresponse conditions. High current responsivities are recorded as the applied bias approaching $V_T$. The illuminations are provided by various continuous-wave lasers at wavelengths, from Vis to THz, λ= 532 nm, 635 nm, 1064 nm, 1550 nm, 10 μm, 96.3 μm, 118.8 μm and 163 μm respectively.

## Noise characteristics of 1T-TaS$_2$ devices.

The NEP is estimated by NEP = $I_n$ / $R_{fm}$, where $I_n$ is the noise current in A·Hz$^{-1/2}$, $R_{fm}$ the responsivity in unit of AW$^{-1}$, and NEP in unit of W·Hz$^{-1/2}$. For evaluating NEP values, the noise current is extracted from the noise power spectra at frequency bandwidth of 50 Hz, and the photoresponsivities are estimated at the same frequency.

The noise power spectra are measured using home-made amplifiers and then digitized with a data acquisition card. A cross correlation algorithm is used to average out the instrument noise. Just battery and ballast resistors were used to avoid extra noises due to external circuit. The device is kept in a shielded dark enclosure.

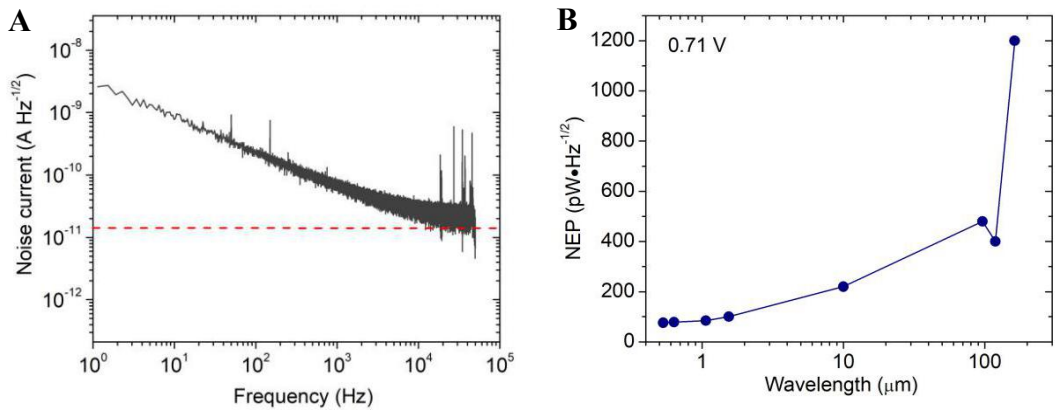

**Fig. S8:** (A), Noise current as a function of frequency for external bias 0.71 V. The red dashed line is guidline for the Johnson–Nyquist noise limit. (B), The extracted NEPs at various incident wavelengths from visible to THz spectral range at frequency bandwidth of 50 Hz. For NEP analysis, the noise power spectra were measured using home-made amplifiers and then digitized with a data acquisition card. A cross correlation algorithm is used to average out the instrument noise. Just battery and ballast resistors were used to avoid extra noises due to external circuit. The device is kept in a shielded dark enclosure when experiment.

**Temporal response of the device after the femtosecond pulse excitations.**

For temporal response analyses, pulsed femtosecond lasers at wavelengths λ = 800 nm (100 fs, 1.2 mJ/cm$^2$) and λ = 2.5 um (150 fs, 1 mJ/cm$^2$) with the repetition rate of 1KHz are provided as laser sources. The temporal electrical signals are recorded by an oscilloscope (Agilent Technologies DSO-X 3034A, the bandwidth is 350 MHz).

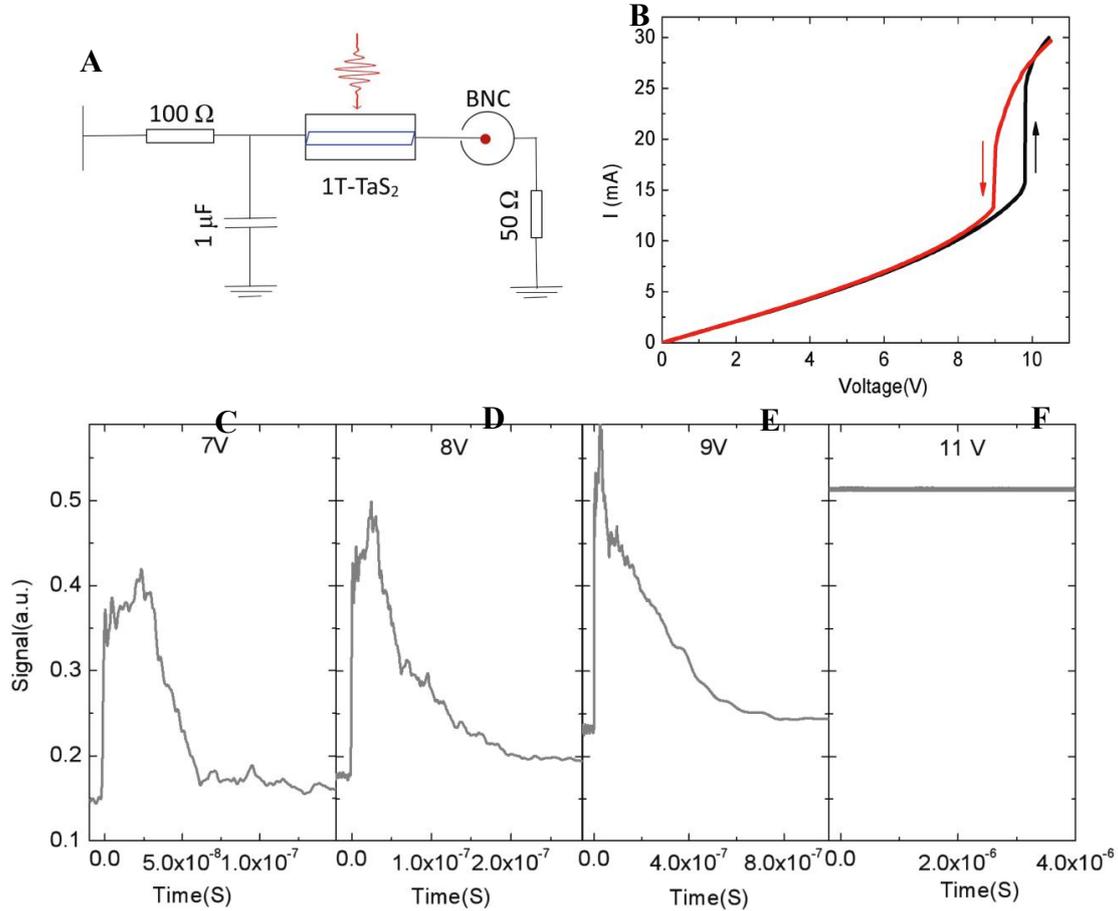

**Fig. S9:** Temporal response of the device after the 150 fs pulse excitation. (A), The measuring circuit. (B), The I-V function of the device under dark for both sweep up and down modes. (C)-(F), The time evolution of the photo-excited signal at various applied biases for an incident intensity (~ 1 mJ/cm$^2$). The laser source is provided by a pulse femtosecond laser at λ = 2.5 μm (150 fs) with a repetition rate of 1 KHz. The temporal 'time-of-flight' response is recorded by an oscilloscope with a bandwidth of 350 MHz, for which a channel in 50 nm thickness and 30 μm length between the electrodes is used for measurement. Here, the contacts are patterned using electron beam photolithography. Given an applied bias below the threshold voltage, as shown in (C)-(E), the photoresponse rise time is recorded to be ~1.5 ns, which is limited by the time resolution of the oscilloscope electronics. After the pulse excitation, the photoexcited signal shows firstly a plateau lasting ~30 ns, then it shows a fast fall in another 30 ns followed by a slow tail. The slow tail is found to be bias-dependent and get slower as the bias increasing. With the applied potential exceeding the threshold voltage, as shown in (F), there is no photoresponse detectable.

**Table.S1. Parameters for 1T-TaS$_2$, graphene and TIs based (THz) photodetectors**

| Description | Device type | Responsivity | Response time (~order) | Spectral range | Reference |
|---|---|---|---|---|---|
| 1T-TaS$_2$ | PC | 0.2 - 4 A·W$^{-1}$ | ~ ns | Vis ~ THz | Our work |
| Graphene stripe | PTE | 8.4 mV·W$^{-1}$ (2.4 μA·W$^{-1}$) | 0.1-15 s | THz | 33 |
| Bi$_2$Te$_3$-Si junction | PC | 1A·W$^{-1}$ (532nm), 1.7 mA·W$^{-1}$ (THz) | ~ 100 ms | UV ~ THz | 23 |
| Graphene THz antenna | PV/Bolometric | < 10 nA·W$^{-1}$ | ~ ns | MIR-THz | 34 |
| Bilayer-graphene FET | PV/PC | 1.2 V·W$^{-1}$ (PV), 1.3 mA·W$^{-1}$ (PC) | - | THz | 35 |
| Graphene FET | PV | 100 mV·W$^{-1}$ | ~ ms | THz | 36 |
| Graphene (gated) | PTE | 10 V·W$^{-1}$ | ~ ps | THz | 37 |
| Graphene FET | Plasmonic/Bolometric | < 1mV·W$^{-1}$ | - | THz | 38 |

\* TIs: topological insulators, PC: photo-current, PV: photo-voltaic, PTE: photo-thermo-electric, FET: Field-effect transistor, UV: ultraviolet, Vis: visible, MIR: Mir-infrared.

\* In the table, the Bi$_2$Te$_3$ is belong to TIs.